\documentclass[12pt,preprint]{aastex}
\usepackage{graphicx}

\def\kms{\mbox{$\rm km~s^{-1}$}}
\def\sec{$^{\prime\prime}$~}
\def\deg{$^\circ$}

\def\arcsec{^{\prime\prime}}

\begin{document}

\title{BIMA MILLIMETER-WAVE OBSERVATIONS OF THE CORE-JET AND MOLECULAR GAS IN THE FR~I RADIO GALAXY NGC 3801}

\author{Mousumi Das\altaffilmark{1}, 
Stuart N. Vogel\altaffilmark{1}, 
Gijs A. Verdoes Kleijn\altaffilmark{2}, Christopher P. O'Dea\altaffilmark{3, 4}, 
Stefi A. Baum\altaffilmark{3, 5}}

\affil{\altaffilmark{1}Department of Astronomy, University of Maryland,
College Park, MD 20742 }
\affil{\altaffilmark{2} European Southern Observatory, Karl-Scwarzschild-Strasse-2,
D-85748, Garching, Germany }
\affil{\altaffilmark{3}Space Telescope Institute, 3700 San Martin Drive,
Baltimore, MD 21218}
\affil{\altaffilmark{4} Department of Physics, Rochester Institute of Technology, 
84 Lomb Memorial Dr., Rochester, NY 14623-5603 }
\affil{\altaffilmark{5} Center for Imaging Science, Rochester Institute of Technology, 54 Lomb Memorial Dr., 
Rochester, NY 14623-5603 }

\email{email : mousumi@astro.umd.edu}

\begin{abstract}
  
  We present BIMA 3~mm observations of the radio continuum source and
  the molecular gas disk in the radio loud Fanaroff \& Riley Type I 
  (FR~I) galaxy 
  NGC 3801. We have detected a continuum source in the nucleus and
  determined that it has a flat millimeter-wave spectrum, suggesting
  that the emission is non-thermal and due to an AGN; the radio core
  is not evident in existing VLA observations.  We also map the extended 3
  mm emission from the previously known radio jets.  In addition, we
  detect CO (1--0) emission associated with the dust disk
  observed in previous HST images.  A velocity gradient is observed,
  indicating a two kpc radius rotating gas ring or disk oriented
  roughly perpendicular to the radio jets.  The inferred molecular gas
  mass of the disk is $M(H_{2})=3\times10^{8}M_{\odot}$, about 1\%\ of
  the dynamical mass.  We also find a $\sim 10^8$~M$_\odot$ molecular
  gas clump not associated with the gas disk.  There is evidence that 
  this gas is associated with a merger and is infalling.  This
  suggests that FR~I type activity is related to merger activity, as is
  thought to be the case for FR~II type radio galaxies.  We also find
  indications that one of the radio jets is entraining gas from the
  infalling molecular gas.
 
\end{abstract}

\keywords{galaxies:active --- galaxies:individual (NGC 3801)--- 
galaxies:ISM  --- interstellar:molecules ---  interstellar:kinematics and dynamics 
---radio lines:galaxies}

\section{INTRODUCTION}

Radio loud galaxies are among the most energetic sources in the nearby
universe. Associated with luminous early type (E/S0) galaxies, they
often have relativistic jets originating in the nucleus near a
supermassive black hole (SMBH) thought to power the enourmous energy
output (eg. Blandford 1984).  Depending upon the jet morphology, the
galaxies are classified as FR~I (edge darkened) or FR~II (edge
brightened) sources (Fanaroff \& Riley 1974). The core and jet
structures have been mapped at various radio frequencies (e.g. Xu et
al. 2000; Giovannini et al. 2005), 
from a few hundred MHz to several GHz.  But radio continuum
observations at millimeter wavelengths ($\sim 100$ GHz) are relatively
rare (Wright \& Birkinshaw 1984; Salter et al. 1989; Evans et al.
1999a; Looney \& Hardcastle 2000; Hardcastle \& Looney 2001).  

Molecular gas has been detected at millimeter wavelengths in several
nearby radio loud galaxies (Lim et al. 2003; Leon et al. 2003; Lim et
al. 2000; Evans et al. 1999a, 1999b; Mazzarella et al. 1993; Mirabel,
Sanders \& Kazes 1989; Philips et al. 1987); the mass of detected
molecular gas is in the range $\sim$10$^{8}$--10$^{10}~M_{\odot}$.
The molecular gas mass is usually larger than that seen in radio quiet
or normal ellipticals, where the gas mass appears in the range
$\sim$10$^{8}$--10$^{9}~M_{\odot}$ (Lees et al. 1991; Wiklind, Combes
\& Henkel 1995; Knapp \& Rupen 1996; Young 2002).  Higher angular
resolution observations which map and resolve the molecular gas
distribution exist for only a few radio galaxies so far: Centaurus A
(Eckart et al. 1990), 3C 84 (Inoue et al. 1996), 
3C 293 (Evans et al. 1999a), PKS 1345+12 (Evans et al. 1999b)
and 3C 31 (Okuda et al. 2005). These observations
indicate that the gas is nearly always concentrated in a disk
associated with the nuclear region. In at least one case (3C 293), the
gas has a disturbed morphology (Evans et al 1999a).  The detection of
large masses of molecular gas in radio galaxies and the location of
this gas in the nucleus suggest that the gas provides a resevoir for
fueling the AGN and powering the relativistically beamed jets.

However, many questions regarding jets in radio galaxies remain. For
example, what is the origin of the gas feeding the AGN, how is the gas
transported into the nuclear region, and what determines the
luminosity of the AGN?  X-ray observations indicate that radio
galaxies can create bubbles and shells within the X-ray emitting gas
in clusters of galaxies (e.g. Heinz, Reynolds \& Begelman 1998; 
Fabian et al. 2003).
Thus jets from radio galaxies may play a significant role in heating
the intracluster gas (see for example Reynolds et al. 2005 and
references therein).  The role of jets may be particularly prominent
in less rich clusters or groups where gravitational heating is not as
strong, and it is therefore of particular interest to observe jets and
their interactions with gas in such systems.
 
To address some of these questions we have undertaken a study of
molecular gas and millimeter continuum emission in several nearby
radio loud galaxies. In this paper we present the first case in this
study, NGC 3801 (UGC 06635).  NGC 3801 is a nearby early type galaxy,
variously classified as an elliptical or an S0 galaxy, located in a
group of approximately ten galaxies; its basic properties are listed
in Table 1.  It is a radio bright galaxy with radio jets originating
from the nucleus approximately normal to a central dust lane.  It is
classified as an FR~I type radio source and has the classic
edge-darkened jet structure typical of its class.  The galaxy has been
extensively imaged with the HST WFPC2 camera by Verdoes Kleijn et
al. (1999). They find a prominent dust lane in the galaxy center with
several associated filamentary structures; there are several knots of
$\rm{H}\alpha$ emission along the main dust lane that likely trace
star forming regions. The radio jets have been mapped with the VLA at
1.5~GHz (Wrobel \& Condon 1989); they extend out to a distance of
$\sim20\arcsec$\ from the nucleus.  The radio core is not detected
with the VLA. HI has been detected toward NGC 3801 using the Arecibo
radio telescope, with a total estimated HI mass of
$2.1\times10^{9}~M_{\odot}$ (Heckman et al. 1983; Duprie \& Schneider
1996). Leon et al.  (2003) detected molecular gas in the center of NGC
3801; their single dish CO observations show that the line profile has
a horn-shaped structure indicative of gas rotating in a disk or
torus. They estimate a molecular gas mass of
$\sim3.8\times10^{8}~M_{\odot}$ in their 21\sec beam. We adopt a
distance of $D=47.9$~Mpc for NGC 3801 ($V_{sys}=3451~\kms$), which 
leads to a scale of 0.23~kpc arcsec$^{-1}$.

In this paper we present a map of the molecular gas distribution in
NGC 3801 using the BIMA millimeter-wave array.  We have also detected
the compact radio core in the center of the galaxy, which was not
detected in lower frequency VLA observations.  We have investigated
the continuum emission at two other frequencies in the 3~mm window to
determine whether the core emission has a flat or steep spectrum. This
will help to understand whether the continuum emission is from an AGN.

\section{OBSERVATIONS, DATA REDUCTION, AND MAPS}

We observed NGC 3801 with the BIMA interferometer (Welch et al. 1996)
in the C and D configurations from 2003 July to 2003 September.  The
observations were all obtained in single-pointing mode with a
100$\arcsec$ field of view centered on the galaxy nucleus.  The 114.2
GHz frequency of redshifted CO (1--0) was positioned in the upper
sideband.  We used the wideband correlator setting, which gives a
velocity resolution of 50~\kms\ and a velocity range of
$\sim$2000~\kms, covering 2100 to 4055~\kms; this setting gives a
lower velocity resolution than narrow band modes, but the large
velocity width enables subtraction of the continuum emission to obtain
a map of CO (1--0) line emission alone. We used the quasar 3C 273 for
flux calibration. The flux of 3C 273 is monitored regularly with BIMA
and varied by approximately 10\% during the period of observation;
additional errors are introduced when the flux calibration is
transferred to the source. So the total error in the flux estimate is
of the order of 10 to 20\%.  The radio source 1159+292 was used for
phase calibration.  Emission in the wideband channels is sufficiently
bright that it could be used to self-calibrate the instrumental
phases; self-calibration produced modest improvement in the images,
and was used for the maps produced here.  All interferometric data
obtained with BIMA were reduced using MIRIAD (Sault, Teuben \& Wright
1995).

\subsection{3 mm Continuum Emission and Spectral Index Maps}\label{contin_obs}

To obtain the continuum emission map, we flagged channels containing
CO (1--0) line emission and recomputed the continuum channels from the
narrow band dataset.  We then averaged the upper sideband
$(\sim$114.3~GHz) and the lower sideband ($\sim$110.6~GHz). The mean
frequency of the continuum emission map is thus 112.4~GHz,
corresponding to a wavelength of 2.7 mm.  Figure~1
shows two maps of the distribution of 2.7 mm continuum emission in NGC
3801.  Both maps combine data from the C and D arrays, but the map in
the left panel is optimized for maximum sensitivity and flux recovery
(using natural weighting) and the other compromises sensitivity
slightly to achieve better resolution (using robust=0.5 weighting).
The synthesized beam full width at half maximum diameters are
$12\times 10\arcsec$ and $10\times 8\arcsec,$ respectively.

To derive the spectral index of mm-wave continuum emission, we carried
out additional BIMA observations at two new frequencies, 86 and
110~GHz.  Although the 110~GHz frequency is not very different from
the 112.4 GHz frequency used for the maps shown in this paper, we
decided to observe two frequencies simultaneously so as to minimize
the effect of calibration uncertainties, including those resulting
from flux variations of the quasar used for flux calibration.  We
observed at 110 GHz rather than 112.4 GHz because the sensitivity
improves away from the atmospheric oxygen band just beyond the CO
(1--0) line.  The spectral index observations were obtained in 2003
November using the C array in single-pointing mode with the wideband
correlator setting.  In order to use the same beam size for all
frequencies, in making the spectral index maps we restricted the uv
range of the visibilities to the same range, 2 k$\lambda$ to 31
k$\lambda$.  We used natural weighting. Even with the uv range
restriction, the beam sizes were slightly different, and so we
convolved all maps to the lowest resolution,
$13.0\arcsec\times7.5\arcsec$.

Also, we used VLA 20 cm (1.5~GHz) archival data and our BIMA 2.7 mm
map to determine the spectral index between millimeter and centimeter
wavelengths.  The highest resolution VLA map has a resolution of
1.5$\arcsec$\ and is shown in the upper left of Figure 2
and was obtained by Wrobel \&\ Condon (1989). The faint diagonal
striping is probably an artifact of incomplete uv coverage in the A
array; the bend in both jets is not an artifact.  In the upper right,
we show a VLA B array 20 cm map at 4$\arcsec$\ emission that has been
convolved to match the 10$\times 8\arcsec$ of the BIMA 2.7 mm map
shown in Figure 1b.

\subsection{CO (1--0) Emission}

For the CO line emitting gas distribution, we need to subtract the
continuum emission.  As noted earlier, CO was observed in the upper
sideband.  The continuum maps presented in Figure 1
contain emission from the lower side band and so were not used.
Instead, we formed a continuum map from off-line channels (2100 to
2900~\kms) in the upper sideband, and used that for continuum
subtraction.  The velocity-integrated CO (1--0) emission map, formed
by taking the zeroth moment of the emission, is shown in
Figure 3.  Flux from pixels with absolute value less than 9
mJy beam$^{-1}$ (1.1$\sigma$) was not included in the map shown.  We
note that clipping results in some of the flux being excluded from the
map, and so in computing fluxes we used unclipped maps.  The
resolution of the natural-weighted map is $12\arcsec\times10\arcsec$.

\section{RESULTS}

\subsection{Radio Continuum Core and Jet Emission}\label{radio_contin}

The distribution of 2.7 mm (114 GHz) continuum emission is shown in
Figure 1.  As described in Section~\ref{contin_obs},
the right panel shows a robust-weighted map ($10\times 8\arcsec$),
while the left panel shows a natural-weighted map with slightly lower
resolution ($12\times 10\arcsec$) but with better sensitivity.  The
total 2.7 mm flux in the natural-weighted map is 51 mJy.  The maps
reveal that the source has both an unresolved core and extended structure.

In Figure 4, the natural-weighted 2.7 mm map (contours)
is overlaid on the A-array VLA 20 cm (1.5 GHz) image.  Here it can be
seen that the 2.7 mm extended emission is similar in position angle
(PA $\sim$ 120\deg) and extent (radius $\sim 15\arcsec$) to the
jet-like distribution evident in the 20 cm data.  Although there is
also 20 cm emission at the position of the 2.7 mm core, it is possible
that this is entirely due to the jets.  A radio core source has not
previously been noted in NGC 3801; in particular, Xu et al. (2000) used
VLA data taken by Wrobel and Condon to set an upper limit of 0.8 mJy
on the 20 cm flux of the core source.

First we discuss the extended emission.  Xu et al. (2000) classified
NGC 3801 as a twin jet source on the basis of the centimeter wave
VLA data, and noted that twin jet or core-jet structures are common in
FR~I radio galaxies.  However, at millimeter wavelengths jets have
only been detected in a few radio galaxies (e.g. Cygnus A: Wright \&
Birkinshaw 1984; 3C20: Hardcastle \& Looney 2001; M87: Salter et
al. 1989).  Our detection of a core-jet structure in the FR~I galaxy
NGC 3801 at millimeter wavelengths is thus an important step in
understanding of radio galaxies.

The curvature in the jets at 20 cm is also faintly seen in the BIMA
2.7 mm map. The curvature may indicate that the jets are precessing or
deflected by irregularities in the interstellar medium of the galaxy
(e.g. Merritt \& Ekers 2002; Schmitt et al. 2002; Caproni \& Abraham
2004; Hardee et al. 1994).  In Figure 6, we overlay the
VLA 20 cm map (contours) on the HST V-band image obtained by Verdoes
Kleijn et al (1999).  As noted by Verdoes Kleijn et al., the jets are
at a position angle (PA) of 120\deg, roughly perpendicular to the
disk-like dust filament at PA=24\deg\ passing near the nucleus.  The
figure also shows that the peak emission in the SE jet appears to
coincide with the innermost extent of the long, well defined SE dust
filament.  We discuss this association further in
Section~\ref{co_line}.

In order to determine the cm to mm wave spectral index of the jets, we
convolved the 1.5$\arcsec$ 20 cm VLA map to the
$10 \times 8\arcsec$ resolution of the BIMA 2.7 mm (112.4 GHz) map, as
described in Section 2.  The smoothed 20 cm map is shown in
Figure 2 (upper right); a spectral index map formed from
the ratio of the 2.7 mm map to the 20 cm map is shown in the lower
left of the figure. The spectral index is $-0.77$ at the SE peak of
cm-wave emission and $-0.75$ at the NW peak; it lies close to these
values everywhere except closer to the nucleus where core emission
contaminates jet emission, and near the edges where the signal to
noise is not as good.  The spectral index is similar to those measured
in samples of FR~I sources at much longer wavelengths (e.g. Parma et
al.  1999), and is further discussed in Section 4.

Now we return to the millimeter-wave core source.  As previously
noted, this source was classified as a ``twin-jet'' source, not a
``core-jet'' source, by Xu et al. (2000) on the basis of the existing
1.5$\arcsec$ 20 cm VLA data.  Among their sample of $\sim$20 nearby
low-luminosity FR~I radio galaxies, NGC 3801 has the third highest VLA
20 cm flux from the core-jet region.  Yet it is among three galaxies
in their sample with no detection of a core source, with an upper limit
for the core source of 0.8 mJy, an order of magnitude below the 17
core detections.

At 2.7 mm, the core is very clearly detected, with a S/N of 18 at the
peak position in the natural-weighted map (Figure 1).
Within the errors (estimated at 1$\arcsec$\ for the BIMA map), the
position of peak 2.7 mm emission coincides with the nuclear position
given in the near-IR 2MASS galaxy catalog (Jarrett et al. 2003), as
shown in Figure 1. The flux at the position of the
nucleus in the higher resolution $10\times 8\arcsec$\ map, which
better resolves the core, is 18.0 mJy beam$^{-1}$.  However, there is
likely to be some contamination from jet emission.  One way to
separate unresolved emission from more extended emission is in the
$uv$ plane; however, the signal to noise is not sufficient for this.
Instead we use the 20 cm map, which is completely dominated by jet
emission, to estimate contamination from the jets.  The spectral index
map shows that the ratio of jet emission at 20 cm to that at 2.7 mm is
very close to 27 everywhere (corresponding to a spectral index of
-0.77).  We divided the 20 cm map shown in the
upper right of Figure 2 by this factor and subtracted the
scaled 20 cm map from the 2.7 mm map.  The result is shown in the
lower right of Figure 2.  Here it can be seen that
contamination from jet emission has been cleanly removed.  The slight
extension at a position angle of about 20\deg, perpendicular to the
jets, is probably not real.  The flux in the 2.7 mm map, after
subtracting the jets, is 15 mJy.  Since the upper limit for the core
at 20 cm is 0.8 mJy, we find that the core is at least 19 times
brighter at 2.7 mm, corresponding to a spectral index greater than
0.68.

To learn more about the core source, we can investigate its spectral
index over the range of wavelengths observable within the 3 mm
atmospheric window.  Figure 8 shows the core flux
at several frequencies in the 3 mm window.  Note that as described in
Section 2 these observations were convolved to a common resolution of
$13\times 7.5\arcsec$, somewhat larger than that used for the core
flux estimate at 2.7 mm, and so there is contamination from the jets
at the 20\%\ level.  Also, because the map contains data only from the
C array, some flux is resolved out, resulting in somewhat lower fluxes
than in the D plus C array maps shown in Figure 1.
Figure 8 shows that the 3 mm spectrum is flat to
within the errors.

Figures 1 and 2 (lower right) show that the
2.7 mm continuum peak appears to lie about 1$\arcsec$ east of the
nuclear position given by the 2MASS near-IR peak (Jarrett et al.
2003); note that the 2MASS position is very close to the peak in the
HST V-band image (Verdoes Kleijn et al. 1999).  The absolute positional
accuracy of interferometers such as BIMA is typically about 10\%\ of
the synthesized beam width.  Since the beam is about 10$\arcsec$, we
expect a positional accuracy of about 1$\arcsec$.  The errors in the
2MASS position should be smaller than this.  Therefore the
millimeter-wave continuum and line observations presented here should
be shifted about 1$\arcsec$ west for accurate registration with the
optical images shown in this paper.

\subsection{CO (1--0) LINE EMISSION}\label{co_line}

The map of CO (1--0) velocity-integrated emission from a region
80$\arcsec$ (18 kpc) on a side centered on NGC 3801 is shown in
Figure 3; Figure 6 (lower panel) shows the
same CO emission contours overlaid on the HST V-band image.  The CO
emission appears distributed in three main clumps, each labeled in
Figure 3.  The HST V-band image shows a straight prominent
dust lane, measured to have a position angle (PA) of 24\deg\ by
Verdoes Kleijn et al. (1999), which is offset approximately
1$\arcsec$\ west of the nucleus.  CO clump A appears associated with
the prominent NE extension of this filament.  CO clump B is roughly
coincident with the SW filament, although it appears offset by a
couple of arc seconds to the east. It is possible that this offset is
simply a result of low signal to noise.  However, we note that the CO
brightnesses of clumps A and B are similar but the SW dust filament is
not as dark; this suggests the clump B gas may be located more towards
the far side of the galaxy.  Also, the major axis of the gaseous disk
passes through the nucleus as might be expected, in contrast to the
dust filament.  We discuss the geometry of the nuclear disk later in
the next section.

Clump C also appears to be associated with a dust filament: it
peaks near the inner end of the well defined east filament that is
oriented perpendicular to the filament associated with clumps A and B.
The signal to noise is too low to determine whether the clump is
extended.

The entire CO (1--0) emitting complex extends over nearly $\sim
30\arcsec$, with the peaks of clump A and B emission separated by
about 20$\arcsec$. Assuming a distance of 47.9~Mpc for NGC 3801, the
molecular gas extends over a region nearly $\sim$7~kpc across, with
the peaks of the A and B clumps separated by approximately 4.6 kpc.
The spectrum of the $30\times 40\arcsec$\ region centered on the
nucleus is shown in Figure 5; emission is detected over a
range of about 600 \kms.

The velocity-integrated CO (1--0) fluxes ($S_{CO}$) of the three
clumps are similar, and are listed in Table 3.  Together, the clumps
total to $S_{CO} = 14.2$ Jy~\kms; this is similar to the flux measured
by Leon et al. (2003) in the 21$\arcsec$\ IRAM 30-m single-dish beam.
Some of the emission we detect is near the half-power response of the
30-m beam, and so the Leon et al. estimate is probably a lower limit
to the total flux.  We note in particular that our spectrum is more
asymmetric than the Leon et al. spectrum.  This might be explained
if the IRAM beam was missing proportionally more of clumps B and C.
However the BIMA map may be underestimating the total flux.  This is
because interferometric estimates of total flux tend to miss weak
emission (see Helfer et al. 2002 and Helfer et al. 2003 for a
discussion of this).  To convert the CO fluxes to molecular mass, we
use the standard conversion factor giving the molecular mass as
$M_{mol} = 1.5\times 10^4 D^2_{Mpc} S_{CO} $ M$_\odot$, where
$D_{Mpc}$ is the distance in Mpc and $S_{CO}$ is the
velocity-integrated CO flux in Jy \kms\ (e.g. Strong et al. 1988;
Scoville et al. 1987).  Note that this formula includes a factor of
1.36 for helium.  We thus find that the three clumps, A, B, C, have
masses 1.3, 1.7 and 1.8$\times 10^8$ M$_\odot$ (Table 3); since the
interferometer may miss some of the flux, these estimates should
probably be considered lower limits.

Figure 9 shows CO position-velocity contour plots for two cuts
through the nucleus.  The upper panel shows a cut at PA=24\deg,
parallel to the dust lane passing near the nucleus.  This cut passes
near the peaks of clumps A and B, which are labeled in the figure.
Clump A has emission at velocities from 3180 to 3340 \kms; clump B
emits from 3580 to 3750 \kms.  The figure shows a velocity jump
between clumps A and B and a velocity gradient within clump B,
suggesting rotation.  We discuss the evidence that this is a rotating
nuclear disk in Section 4.2.  The lower panel shows a cut
perpendicular to the first (PA=114\deg), also through the nucleus.
Clump C appears in this cut, with emission in the range 3470 to 3630
\kms.

To interpret these velocities, we need to know the systemic velocity
of the galaxy.  Values given in the literature for the redshift of NGC
3801 show a wide range; NED lists seven redshifts, in the range 3230
to 3460 \kms.  However, not all of these are of similar quality.  The
Arecibo spectrum in DuPrie and Schneider (1996) is well determined
with a good baseline.  They report a systemic velocity of 3451 \kms.
Their spectrum also shows absorption in the profile. The absorption,
spectral baselines of limited extent, and confusion from other
galaxies in the group may have confused some earlier HI observations.
Given the fact that some of the HI may be due to merging gas and
that existing HI observations do not resolve the galaxy, it is
important to have confirming estimates of the systemic velocity
using stellar tracers.  A recent determination by Wegner et
al. (2003), based on stellar absorption lines using a
cross-correlation technique, gives a velocity of 3494 \kms, with an
error of 70 \kms.  This should avoid the uncertainties resulting from
merging or asymmetrically distributed gas, since this 
determination is based on stars near the galaxy center.  Similarly,
the redshift for the Updated Zwicky Catalog used for the CfA survey is
3456 \kms\ with an error of 71 \kms.  We adopt a systemic velocity of
3451 \kms, the estimate from the Arecibo HI spectrum.

The velocity extent of CO emission is very similar to that of the HI
21 cm profile observed by DuPrie and Schneider (1996).  Also, the HI
spectrum is in absorption in the velocity range 3460 to 3600 \kms,
which corresponds closely to the velocity range of clump C 
(approximately 3470 to 3630 \kms).  As mentioned earlier, the C
component coincides with the inner part of the eastern dust filament
(Figure 6); the large extinction implies that the dust
filament and therefore the CO associated with it is on the near side
of NGC 3801.  We note also that this gas is along the line of sight to
the SE radio continuum jet, as seen in Figures 6 and
7. As is also evident in the figures, the 20 cm radio
continuum comes almost entirely from the twin radio jets; therefore
the 21 cm absorption must be produced by gas on the line of sight to
the jets.  This is very likely to be HI gas in the eastern dust
filament, given its overlap with the SE radio jet.  The fact that the
HI absorption and clump C velocities are similar further increases
the likelihood of this association, since the clump C coincides with
the filament.  We therefore conclude that the clump C is associated
with the east dust filament, which is on the near side of NGC 3801,
and that this filament produces the absorption seen in HI.  Although
the similarity in the position of clump C with the peak of the
SE radio jet suggests a physical association, it is unclear whether
this is the case.  We discuss this further in the next section.

\section{DISCUSSION}

\subsection{Nature of the AGN Core and Jets}

Our BIMA detection of $\sim$3 mm continuum emission from
the core and jets in NGC 3801 is one of the key results of this paper,
because radio cores and lobes have been detected at millimeter
wavelengths in so few galaxies to date. Our 3 mm continuum
observations reveal a bright radio core (Figure~1)
which was not detected in 20 cm VLA observations
(Figure~4). In the sample of 17 FR~I galaxies studied by
Xu et al. (2000), NGC 3801 was the only galaxy that lacked a VLA core
and was hence not observed with the VLBA.  One possible explanation
for the absence of a core source is Doppler deboosting, which occurs
for relativistic jets in the plane of the sky; the jets in NGC 3801
are probably close to the plane of the sky, given the similarity in
fluxes from the two jets even close to the source and also the edge-on
orientation of the nuclear disk.  However, our BIMA observations 
clearly show that NGC 3801 has a bright radio core at 3 mm.  This
implies that the weakness of centimeter-wave emission from the core cannot be
attributed to deboosting.  Instead it is likely that the core is
compact and dense.  Synchrotron self absorption scales as ${\nu}^{-3}$
(e.g. Krolik 1999) and can explain the absence of centimeter-wave emission.
At higher frequencies such as the 86--114 GHz frequencies observed
here, the compact core is optically thin and thus appears
bright. Synchrotron self absorption has also been observed in other
radio cores (e.g.  Rudnick, Jones \& Fiedler 1986).  Free-free
absorption is also a possibility, but so far it is rarely seen in
radio galaxies and even then seems to absorb only part of the inner
regions (e.g., Walker et al. 2000; Jones et al. 2001).  However, the
absence of optical emission suggests that free-free absorption is also
a plausible mechanism to explain the absence of centimeter wave
emission from the core.

We find that the core spectrum is flat in the 3 mm window
(86--114~GHz) (Figure~8 and Table~2).
Using the error limits on the peak fluxes at 86~GHz and 112.4~GHz, the
range for the spectral index could however be -0.1 to -0.9. But after including
the emission at 110~GHz, Figure 8 shows the core spectrum
to be nearly flat.  Steeply
positive spectral indices, such as produced by dust emission
(e.g. Wolfire \& Churchwell 1994), are excluded.  Optically thin
free-free emission, such as might be produced by H II regions in a
starburst, could produce a flat spectrum.  However, the upper limit of
0.8 mJy on the 20 cm core flux rules out this possibility.  The
continuum flux from nuclei of starburst galaxies becomes higher with
increasing wavelength at wavelengths greater than 2 cm (due to
non-thermal emission), but shortward of 2 cm free-emission is
optically thin (e.g. Turner and Ho (1983)).  This means that the
free-contribution at 3 mm must be well below 0.8 mJy, which is much
less than the observed 15 mJy.  We therefore conclude that the 3 mm
continuum emission cannot be attributed to a starburst.
Since flat millimeter-wave spectrum cores are often produced by
non-thermal emission in other types of AGN than those in FR~I
galaxies, we conclude that the core emission is likely produced by an
AGN.  The flatness of the spectrum implies that we are seeing a compact
source which is optically thick at 3mm. However the source may have
a range of size scales, like a jet with some opening angle
so that different parts of the structure become
optically thin at different wavelengths (e.g., Blandford \& Konigl 1979;
Marscher 1980; Ghisellini \& Maraschi 1989).

The bright, compact core and twin jet structure of NGC
3801 lend support to the idea that FR~I sources are the parent
population of BL Lac objects (Trussoni et al. 2003; Urry \& Padovani
1995).  According to this hypothesis, BL Lac radio sources are FR~I
nuclei in which the line of sight coincides with the axis of the radio
jets; the resultant relativistic beaming effect produces very luminous
radio cores. In NGC 3801, the twin jet structure indicates a large
beaming angle. But it was puzzling that the radio core was not
visible; our BIMA results confirm that there is a bright radio core at
the galaxy center. Our observation of the flat spectrum core in NGC
3801 is thus consistent with the unification scheme for AGNs in radio
loud galaxies.

\subsection{The Central Molecular Gas Disk}

As described in Section~\ref{co_line}, CO clumps A and B appear
associated with the dust filament at a PA of 24\deg\ that is offset
about 1$\arcsec$\ west of the nucleus.  This filament resembles a disk
seen nearly edge on (see Verdoes Kleijn et al 1999); if we take the
distance of the peaks of clumps A and B from the nucleus ($\sim
10\arcsec$) as the radius of the disk, then the fact that the filament
is offset $\sim 1\arcsec$\ west of the nucleus suggests a disk with a
10/1 axis ratio, implying a disk with an inclination of 84\deg\ with
the near side facing east.  The position-velocity cut at $PA=24$\deg\
(Figure~9) is also consistent with a disk seen close to edge
on.  Both clumps A and B are offset from the nucleus by about 250
\kms, assuming a systemic velocity of of 3451 \kms.  The velocity
gradient for the 2 kpc CO disk is in the same direction as the
gradient seen in H$\alpha $ and [NII] on scales of a few hundred pc in
STIS HST data by Verdoes Kleijn et al. (1999).  If we interpret the
velocity gradient as rotation, then with a nearly edge-on inclination
of 84\deg\ the circular velocity is $v_c = 250$ \kms\ at a radius of
2.3 kpc (10$\arcsec$), which gives a rough estimate of the dynamical
mass $M_{dyn}=v_{c}^{2}r/G\approx 3\times10^{10}$~M$_{\odot}$.  The
combined molecular mass of CO clumps A and B is about $3\times 10^8$
M$_\odot$, about 1\%\ of the dynamical mass within this radius.  The
HI seen in emission is at similar velocities to CO clumps A and B; if
we attribute the HI emission to the same regions, then the mass in
neutral atomic gas is $1.2\times 10^9$ M$_\odot$, about four times the
molecular mass, and the total gas mass is about 5\%\ of the dynamical
mass.

The total molecular gas mass in NGC 3801 is lower than in most spiral
galaxies but large compared to gas poor ellipticals.  The molecular
gas surface density in NGC 3801 averaged over the $R_{25}$ radius is
about 0.2 M$_\odot$ pc$^{-2}$, similarly small compared to most
spirals (e.g. Sheth et al. 2002).  For the molecular gas disk itself
(e.g. clumps A and B), adopting a radius of 2.3 kpc the average
face-on surface density is 18.4 M$_\odot$ pc$^{-2}$; this is large
compared to the molecular surface density in the solar neighborhood
but somewhat smaller than the average for the Milky Way at a radius of
5 kpc (i.e. including the 5 kpc ring).

We can place constraints on the star formation rate in NGC 3801 using
the infrared luminosity of $2.4\times 10^9$ L$_\odot$ derived from the
IRAS 60 and 100 $\mu$m fluxes (Knapp et al. 1989) using the relations
given by Cox (1999).  Using the relation $SFR=4.5\times10^{-44}~L_{FIR}$
(Kennicutt 1998) to derive the star formation rate (SFR), we obtain a
SFR$\sim$0.4~M$_{\odot}$yr$^{-1}$.  We can also estimate the star
formation rate using the H$\alpha$ $+$ [NII] luminosity observed with
HST by Verdoes Kleijn et al. (1999).  Using the star formation
relation for H$\alpha$ in Kennicutt (1998)) and assuming all the
luminosity is from H$\alpha$, we find a star formation rate of 0.2
M$_\odot$ yr$^{-1}$, a factor of two lower than the infrared estimate.
It is possible that the optical estimate is affected by dust
extinction in the edge-on disk.  This is a fairly low SFR compared to the
disks of nearby spiral galaxies, which have a SFR of a few
M$_{\odot}$yr$^{-1}$, but larger than gas poor ellipticals.  

By contrast, the other FR~I mapped at high resolution, 3C 31, has a
more massive $10^9$ M$_\odot$ disk with a radius of about 1 kpc
(Okuda et al. 2005).  This yields a surface density of 300M$_\odot$
pc$^{-2}$, larger even than the median for the central 500 pc radius
of nearby spiral galaxies (Sakamoto et al. 1999; Sheth et al 2002).
Yet Okuda et al. (2005) note the absence of detected star formation in
3C 31, very different from NGC 3801.

Assuming a present rate of $SFR\sim$0.4~M$_{\odot}$yr$^{-1}$, the
molecular gas will last for $\sim 10^{9}$~yr before being exhausted by
star formation.  Including the HI reservoir, the gas can last for well
over $10^9$ yr.

\subsection{Clump C: Infalling Molecular Gas}

The nature of the eastern CO clump, clump C, is particularly
intriguing.  It does not appear to be associated with the central gas
disk.  If it were in that disk, because it is near the minor axis its
deprojected distance from the nucleus would be 20 kpc, an order of
magnitude further than clumps A and B.  Also, it has emission in the
range 3470 to 3630 \kms, significantly redshifted relative to the 3451
\kms\ systemic velocity; given that it would be on the disk minor
axis, this velocity is highly non-circular. Assuming the disk
orientation described in Section~\ref{co_line}, it would be outflowing
at an improbably large velocity.  We conclude that it is not in the
central gas disk.

Another possibility is that clump C is part of a second disk oriented
perpendicular to the central disk.  NGC 3801 is sometimes classified
as an S0 galaxy, and the east filament partly associated with clump C
might be the the dusty S0 disk seen edge on.  However, the odds of
encountering the specific geometry and orientations require are small,
roughly one in a thousand.  In detail, the ``SO disk'' would have to
be edge-on to the line of sight to within less than 10\deg\ to explain
the thinness of the filament, 2) the nuclear edge-on disk would have
to be within 10\deg\ of edge-on and 3) the two disks would have to be
perpendicular to each other to within 10\deg.  More importantly,
closer examination suggests that NGC 3801 is not an S0 galaxy: the
isophotes beyond a radius of 30$\arcsec$\ are boxy (e.g. Heckman et
al. 1999), consistent with a recent merger, and the inner isophotes
show no evidence for a stellar disk that would be associated with an
S0.  Finally, we note that the gas in the two disks are on colliding
orbits.  If the eastern filament is an S0 disk it must have existed
for some time.  Yet the $PA=24$\deg\ disk seems well established.  Two
disks with intersecting orbits cannot survive.

A third unlikely scenario is that the clump C gas was entrained from
the central gas disk by the eastern AGN jet.  This seems unreasonable
first because the mass of clump C is comparable to the mass of disk
itself; it is difficult to understand how such a large fraction of the
disk could pass in the path of the jet.  Also, at the end of
Section~\ref{co_line} we presented strong evidence that the clump C is
on the near side of the galaxy; in that case, if clump C is entrained
from the central disk it should be blueshifted, contrary to our
observations.

The most likely scenario is that clump C is gas resulting from a
recent merger.  Other evidence for a recent merger exists, such as the
boxy stellar isophotes at radii larger than 30$\arcsec$\ (Heckman et
al. 1996) and the warp in the $PA = 24$\deg\ dust disk evident in the
HST V-band image of Verdoes Kleijn et al. (1999).  As previously
noted, clump C is associated with the eastern dust filament; the shape
of this filament is consistent with the orbital path of tidally
disrupted material.  Molecular emission is only detected associated
with the inner part of the filament. This may be because that is where the
largest gas mass exists, or perhaps because the gas undergoes a
transition to molecular form in the higher ISM pressure inner galaxy.

Since clump C must be on the near side of the galaxy and it is
redshifted relative to the galaxy, we conclude that it is falling
towards the galaxy.  Its projected distance from the nucleus is only
two or three kpc, and so it should soon interact with the 2 kpc
nuclear disk.

\subsection{The Role of Mergers in  FR~I and FR~II Activity}

There is considerable evidence that mergers fuel FR~II radio sources,
which are more energetic than FR~I objects such as NGC 3801.  Baum,
Heckman, and van Breugel (1992) studied a sample of 40 radio galaxies.
Using optical emission line studies, they concluded that all the FR~II
galaxies showed evidence for dynamically young gas disks acquired via
mergers, and suggested that this plays a key role in the FR~II
phenomenon.  There is also evidence for gas disks with abundant
molecular gas in FR~II sources.  In particular, Evans et al. (1999)
found that the FR~II galaxy 3C 293 has an asymmetric gas disk with a
molecular hydrogen mass of $1.5\times 10^{10}$ M$_\odot$.  The gas is
distributed over a 2.8 kpc radius region, and may constitute as much
as 10\%\ of the dynamical mass of the galaxy within that radius.

For the less energetic FR~I galaxies, the suggestion has been that the
AGN may be fueled by gas acquired from the intracluster medium via a
cooling flow (e.g. Baum et al. 1992), motivated in part by the
tendency for FR~I galaxies to be found in clusters.  However, our
observations together with the recent Okuda et al. (2005) Nobeyama map
of the FR~I galaxy 3C 31 
suggests that mergers in addition to cooling flow gas, is  
important for powering the AGN.  The gas disk in NGC 3801 is three
times less massive and has a surface density nearly a factor of 20
times lower.  But it is nonetheless a signficant amount of gas.
Perhaps more significantly, in NGC 3801 there is abundant evidence
that gas has recently arrived via a merger (e.g. clump C) and that the
gravitational effects of the merger are significant (e.g. the warp in
the dust disk, the asymmetric distribution of gas in the disk, and the
boxy optical isophotes).  Hence, it appears that mergers and the
resultant gas inflow (especially gas in non-coplanar orbits) may also
play a key role in the FR~I phenomenon.  Based on the limited high
resolution studies so far, the gas masses (and the gas fraction
relative to the dynamical mass) seem somewhat lower in FR~I galaxies
compared to FR~II galaxies.  However, observations of a larger sample
of galaxies and at higher angular resolution will be essential to
understanding gas fueling in radio galaxies; such observations are
expected soon with the CARMA millimeter-wave array.

\subsection{Jet Entrainment of Molecular Gas?}

Figure~7 shows that the projected eastern radio jet
passes almost directly through clump C.  This raises the question
of whether the jet is in fact interacting with the molecular cloud or instead
simply appears projected behind the CO cloud.  Powerful supersonic
jets, such as those found in FR~II galaxies, overtake ambient clouds
and inflate an overpressured cocoon in which the gas expands mostly
perpendicular to the jet axis, driving a bow shock into the ambient
gas (e.g. O'Dea et al.  2002; 2004).  However, such expansion is not
seen in clump C.  The HI gas, which is seen in absorption and as
previously discussed must be on the near side of the jet, has nearly
the same velocity range as the CO gas.
The similarity in velocity range argues against
expansion such as produced by highly supersonic FR~II jets (e.g. O'Dea
et al. 2002, 2004). Thus the jet is not affecting the gross kinematics 
of the cloud.

A more likely possibility is jet entrainment (e.g. De Young 1986;
Bicknell 1986).  Note that earlier we ruled out the possibility that
clump C is gas entrained from the nuclear disk; here we examine the
possibility that the jet is entraining gas from clump C.  Jet
entrainment is thought to occur in transonic jets, such as those found
in FR~I galaxies.  Here gas in the turbulent boundary layer along the
edge of the jet is entrained and accelerated mainly along the
direction of the jet flow.  In this case, the kinematics of gas on the
near side (e.g.  HI in absorption) and the far side should not show a
systematic difference.  Thus, the velocities associated with clump C
are consistent with entrainment.  In fact, the position-velocity
diagram for clump C shown in the lower panel of Figure~9 shows
possible evidence supporting the entrainment picture.  Note that the
velocity of clump C becomes less redshifted with increasing distance
from the nucleus.  This is in the opposite sense of the gradient
expected for a rising rotation curve.  In principle, it could be
explained as a declining rotation curve.  However, the velocity
decreases from about 3600 \kms\ at 7$\arcsec$ from the nucleus to
about 3540 \kms\ beyond 10$\arcsec$ from the nucleus, much too large a
decrease to be attributed to a declining rotation curve.  However, the
decreasing redshift with distance from the nucleus is in the correct
direction expected for jet entrainment if the eastern jet is the
approaching jet, as supported by our observations.

Figure 4 also shows that the eastern jet appears to bend
just after appearing to pass through clump C, suggesting a possible
association.  We note, however that the opposite jet also bends at the
same distance from the nucleus, but there is no detected molecular gas
concentration.  Such bends can occur due to interaction with the
rotating ISM of a galaxy or due to precession of the jet; we favor the
latter interpretation since the one clear gas rotation axis in NGC
3801 (the $PA = 24$\deg\ disk) has the wrong orientation to attribute
the jet bend to rotation of the ISM.  The jet axis close to the
nuclear disk is offset from normal to the disk by about 6\deg; the
offset is not suprising as the radio jets in nearby radio galaxies are
not always normal to the large scale gas disk (e.g. Wilson, Yang \&
Cecil 2001; Verdoes Kleijn \& de Zeeuw 2004).  The inflow from the
disk that presumably feeds the black hole can produce torques leading
to jet precession.

Also we note the remarkable alignment of the eastern jet with the
eastern dust filament, evident in Figure~6.  The
initial bend in the jet coincides almost exactly with a bend in the
dust filament. Further from the galaxy the jet bends further to the
north while the filament continues nearly in a straight line for more
than 20$\arcsec$\  (4.6 kpc).  This might suggest ballistic motion of the
denser gas while the more diffuse radio gas bends to the north.
However, given that the dense gas was entrained by the jet, it is
difficult to understand how it can separate from the jet.  Also, it is
not obvious how the entrained gas could remain so collimated.  Thus we
prefer our earlier conclusion that the filament traces the orbital
path of the tidally disrupted merging gas of clump C.
 
\section{CONCLUSIONS}  

We have mapped and detected 3 mm continuum and CO (1--0) emission from
the Fanaroff \& Riley type I galaxy NGC 3801 using the BIMA array.  We draw
the following conclusions:

1. We detect 3 mm continuum emission from twin radio jets of similar
brightness and extent and an unresolved nuclear core.  Comparing to a
VLA 20 cm map, we find that the 3 mm and 20 cm jets coincide, and
derive a spectral index of $-0.7$, consistent with synchrotron
emission.  The core has not previously been detected in the radio; we
derive a 3 mm flux of 15 mJy, substantially brighter than the 20 cm
upper limit of 0.8 mJy.  The millimeter-wave spectrum is flat, as
generally observed for the compact radio core of an AGN; other
emission mechanisms appear unlikely.  Our detection of the bright
compact core, along with the twin jet structure, supports the idea
that FR~I galaxies are the parent population of BL Lac objects.

2. We mapped CO (1--0) emission in a region of roughly 100$\arcsec$\
(23 kpc) diameter and detect CO concentrated in three regions, each
with molecular masses in excess of $10^8$ M$_\odot$\ assuming the
standard conversion factor for CO. Unresolved HI emission is seen over
a similar velocity range.  We are able to associate each of the CO
clumps with an HI emission or absorption component.  The molecular
mass is approximately 25\%\ of the HI mass.

3. Two of the CO clumps are associated with a nearly edge-on dust disk
imaged in previous HST V-band observations; the disk is oriented
approximately perpendicular to the twin radio jets.  The CO and dust
distributions suggest an inhomogeneous disk (i.e. not axisymmetric) or
ring with a radius of about 2.3 kpc.  The velocity shift between the
two regions and across one of the clumps is consistent with rotation
reaching a maximum of 250 \kms.  We infer an inclination of 84\deg\
from the offset of the dust major axis from the nucleus relative to
the radius of the disk or ring.  The implied dynamical mass is about
$3\times 10^{10}$ M$_\odot$; thus the molecular mass is approximately
1\%\ of the dynamical mass.

4. We find evidence that the other CO region, clump C, is infalling
gas from a recent merger.  Perhaps most intriguing, there are
suggestions that one of the radio jets is entraining material from the
clump.

5. There are abundant signs that NGC 3801 has undergone a recent minor
merger.  These include the infalling gas, boxy optical isophotes, the
structure of the dust filaments, and the inhomogeneous distribution of
gas.  We therefore suggest that, as in F-R~II galaxies, mergers play a
role in the F-R~I phenomenonon.

\acknowledgments

We thank S.~White for very useful discussions, especially regarding
the VLA data. Conversations with E. Ostriker, C. Reynolds, and J.
Stone were very helpful.  We thank the referee for many very
constructive comments.  Observations with the BIMA millimeter-wave
array are partially supported by NSF AST-0228974.  This research has
made use of the NASA/ IPAC Infrared Science Archive, which is operated
by the Jet Propulsion Laboratory, California Institute of Technology,
under contract with the National Aeronautics and Space Administration.

\clearpage
{\scriptsize
\begin{deluxetable}{lccc}
\tablenum{1}
\tablewidth{0pt}
\tablecaption{Parameters of NGC 3801}
\tablehead{
\colhead{Parameter} & \colhead{Value} & \colhead{Reference}
}
\startdata
Galaxy Type & S0/a & 1 \\
Galaxy Position (RA, DEC) & $11^{h}40^{m}16^{s}.9$, $17^{\circ}43^{\prime}40.5^{\prime\prime}$ & 1 \\
Other Names & UGC 06635, 4C +17.52 & 1 \\
Velocity \& Redshift & 3451~\kms, 0.011 & 2,1 \\
Linear Distance Scale & 0.23~kpc/arcsec & ... \\
PA of Dust Lane & 24\deg$\pm$~2\deg & 3 \\
PA of Radio Jets & 120\deg & 4 \\
\tablerefs{
(1) 2MASS All Sky Survey, (2) Duprie \& Schneider 1996 (3) Verdoes Kleijn et al. 1999,
(4) Xu et al. 2000 }
\enddata
\end{deluxetable}

\clearpage
{\scriptsize
\begin{deluxetable}{lccc}
\tablenum{2}
\tablewidth{0pt}
\tablecaption{Results of Continuum Observations of NGC 3801}\label{contin_results}
\tablehead{
\colhead{Parameter} & \colhead{Value}
}
\startdata
Peak Flux of Core at 86.5~GHz (mJy/beam) & 15.1$\pm$1.6 \\
Peak Flux of Core at 110.2~GHz (mJy/beam) & 10.6$\pm$1.9 \\
Peak Flux of Core at 112.4~GHz (mJy/beam) & 15.5$\pm$1.5 \\
                (beam=$13.0\arcsec\times7.5\arcsec$) \\ 
Position of Radio Core (RA, DEC) & 11:40:16.9, +17:43:41 \\
Flux in Radio Core (mJy) & 15 \\
Flux in Radio Lobes (mJy) & 36 \\
Total Flux (mJy) & 51 \\
P.A. of Radio Lobes & 20\deg \\
Length of Radio Lobes & $\sim$35\sec \\

\enddata
\end{deluxetable}

\clearpage
{\scriptsize
\begin{deluxetable}{lccc}
\tablenum{3}
\tablewidth{0pt}
\tablecaption{Results of CO Observations of NGC 3801}\label{CO_results}
\tablehead{
\colhead{Parameter} & \colhead{Value}
}
\startdata
Total CO Flux (Jy~\kms) & 14.2 \\
~~~~CO Flux of Clump A & 3.8 \\
~~~~CO Flux of Clump B & 5.0 \\
~~~~CO Flux of Clump C & 5.3 \\
CO total line width (\kms) & 500 \\
Rotational velocity of gas disk (\kms) & ~250 \\
P.A. of Gas Disk & 24\deg \\
Total Molecular Gas Mass $M(H_{2})$ ($M_{\odot}$) & $4.9\times10^{8}$ \\
Ratio of molecular to dynamical mass $M(H_{2})/M_{dyn}$ & 0.015 \\

\enddata
\end{deluxetable}

\clearpage

\centerline{\bf Figure Captions}

\noindent
Figure 1.  BIMA maps of 2.7 mm continuum emission in NGC 3801;
  emission is averaged over the upper and lower sidebands, with a mean
  frequency of 112.4~GHz. The ``X'' marks the 2MASS near-infrared
  position for the nucleus.  
  The beam is shown in the lower left of each panel.  (a) The left
  panel shows a map with natural weighting for maximum sensitivity,
  which yields a beam size of $12\arcsec\times10\arcsec$.  The peak is
  20.9 mJy beam$^{-1}$ and the noise level is 1.1 mJy beam$^{-1}$.
  Emission is contoured at 3, 6, 9, 12, 15,
  and 18 times the rms noise level, with negative contours dashed.
  (b) The right panel shows a map with robust = 0.5
  weighting, giving better resolution.  The beam size is $10\arcsec
 \times 8\arcsec$.   The peak is 18.0 mJy beam$^{-1}$ and the noise
  level is 1.4 mJy beam$^{-1}$.  Emission is contoured at 2, 4, 6, 8,
  10, and 12 times the noise level.

\noindent  
Figure 2. (Upper left) Map of 20 cm (1.5 GHz) radio continuum emission from
  NGC 3801 obtained in the VLA A array by Wrobel \&\ Condon (1989).
  The map has a beam size of $1.5\times 1.5\arcsec$ (indicated in
  lower left) and is contoured at intervals of 2, 4, 8, 16, and 32 mJy
  beam$^{-1}$.  The 2MASS near-IR position for the nucleus is
  indicated by the X in this and all panels.  (Upper right) Map of 20
  cm radio continuum obtained in the VLA B array.  The map has been
  convolved to a resolution of $10\times 8\arcsec$ to match the
  resolution of the BIMA 2.7 mm map shown in the right of
  Figure 1.  The map is contoured at intervals of 30,
  60, 120, and 240 mJy beam$^{-1}$.  (Lower left) Spectral index map
  formed from the $10\times 8\arcsec$ 20 cm and 2.7 mm maps is shown
  as greyscale.  The greyscale is a square-root stretch from $-$0.9
  (white) to 0 (black).  The black lines show spectral index contours
  at $-0.8$, $-0.7$, and $-0.6$.  Most of the area of each lobe has a
  spectral index 
  between $-0.8$ and $-0.7$.  The white contours are the same contours
  as shown in the $10\times 8\arcsec$ 20 cm map.  (Lower right) Map of
  2.7 mm continuum emission after subtracting emission from the radio
  jets, as described in the text.  Contours are at 27.5, 55, 82.5,
  110, and 137 mJy beam$^{-1}$.

\noindent
Figure 3. Map of velocity-integrated emission in the CO (1--0) line
in NGC 3801 obtained with BIMA. The contours are at 2, 3, 4, and 5
times the rms noise level, which is 13~Jy~beam$^{-1}$~\kms.
The $12\arcsec\times10\arcsec$\ beam is shown in the
lower left. The 2MASS position for the nucleus  is marked with an X. 
The three
peaks in CO (1--0) emission are marked as A, B, and C. 

\noindent
Figure 4. Contours of BIMA 2.7 mm continuum emission overlaid
  on a greyscale image of VLA 20 cm emission.  The continuum
  emission is contoured as in Figure~1a.

\noindent
Figure 5. BIMA spectrum of CO (1--0) emission in NGC 3801 summed over a
$30\times 40\arcsec$\ box.

\noindent
Figure 6. VLA 20 cm radio continuum (upper panel) and BIMA
  velocity-integrated CO (1--0) emission (lower panel) overlaid on
  HST F555W image of NGC 3801.  The 20 cm and CO (1--0) images are
  contoured as in Figures~2 and 3,
  respectively.  The HST image has been processed to remove the
  radial gradient in light from the host galaxy so as to better reveal
  dust features.  The galaxy nucleus is marked with an X.

\noindent
Figure 7. Contours of BIMA CO (1--0) velocity-integrated emission overlaid
  on a greyscale image of VLA 20 cm emission.  CO emission is
  contoured as in Figure 3.

\noindent
Figure 8. Flux of the radio continuum core in NGC 3801 as a 
  function of frequency
  at millimeter-wavelengths on a log-log plot.  One $\sigma$ error bars
  are shown.  The dotted line shows a $\nu^{3.5}$ power law
  characteristic of dust emission, which is ruled out.  The
  millimeter-wave spectrum can be fitted by a flat spectrum.

\noindent
Figure 9. Position-velocity (PV) plots of CO (1--0) emission for two cuts
passing through the nucleus of NGC 3801.  The upper panel shows the
cut at PA = 24\deg\  corresponding to the major
axis of the nuclear disk; the lower panel shows the perpendicular
direction (PA=114\deg ).  The systemic velocity is
indicated by the dotted line.  Contours are at 14, 21, and 28 mJy
beam$^{-1}$.  The PA=24\deg\  cut passes through CO clumps A and B,
which are labeled.  The PA=114\deg\  cut passes through clump C.

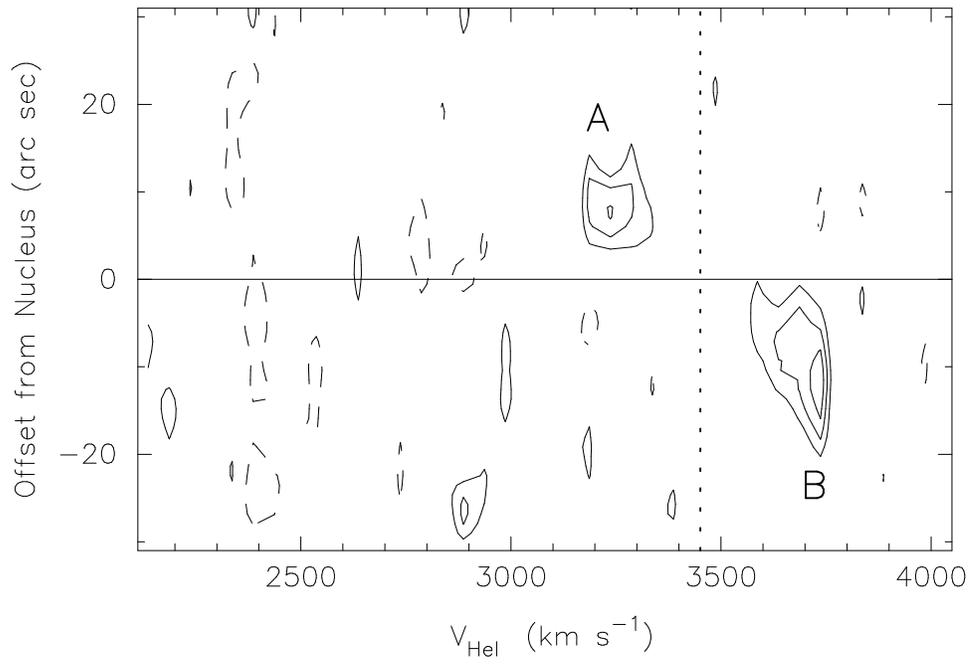
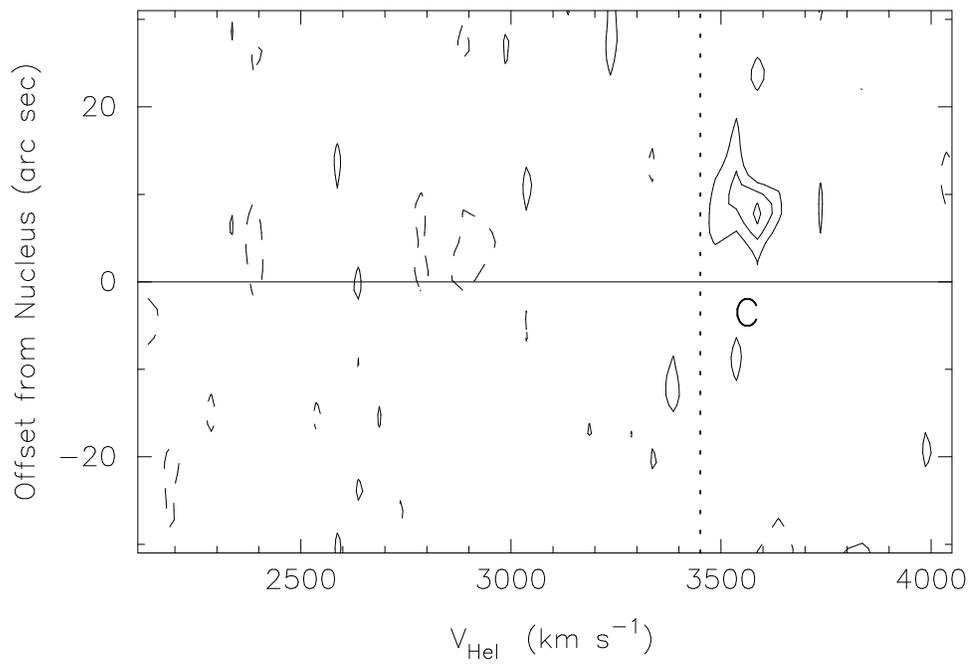
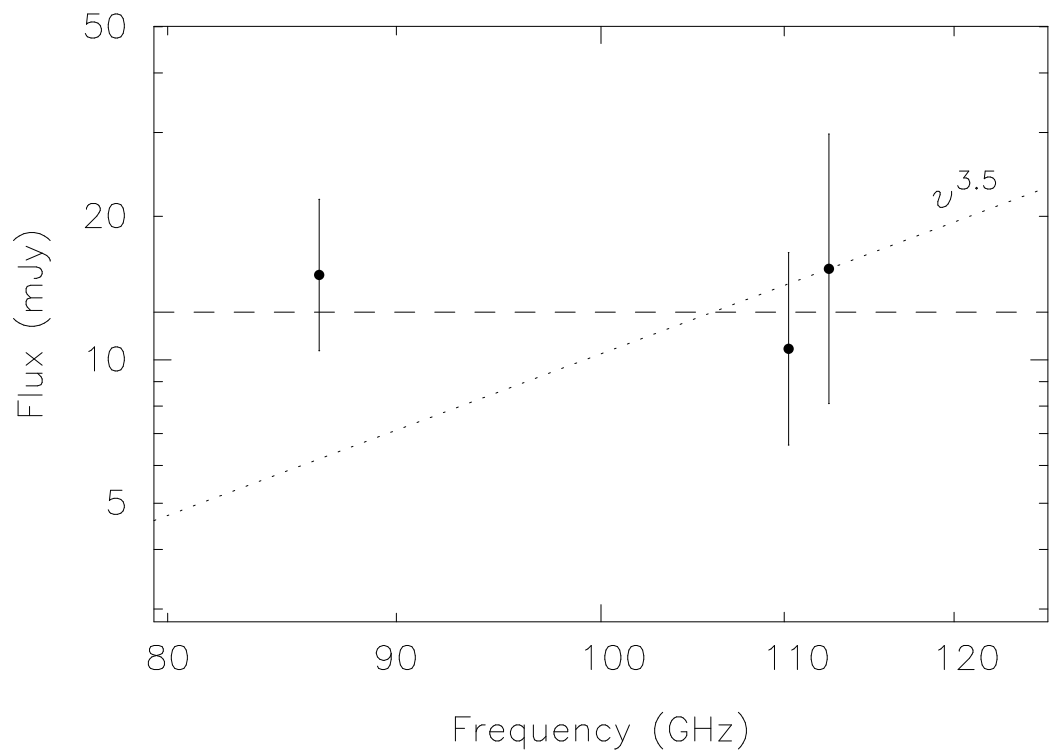
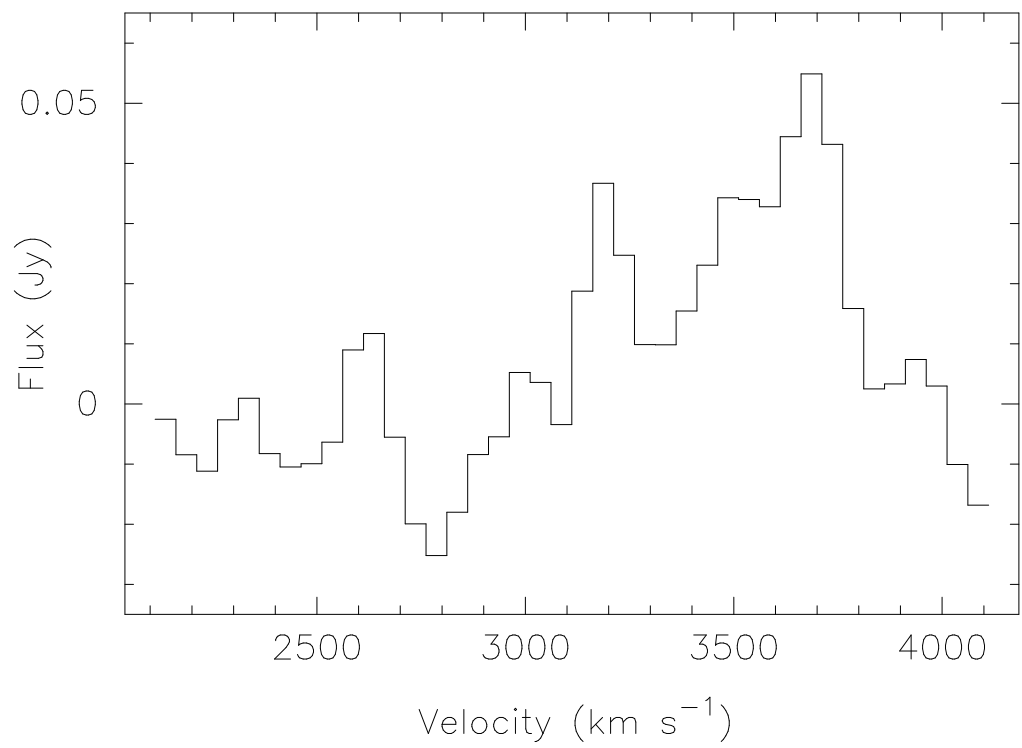
\end{document}